\documentclass[prb,twocolumn,floatfix,showpacs]{revtex4}
\usepackage{graphicx}
\usepackage{times}
\newcommand{\be}{\begin{eqnarray}}
\newcommand{\ee}{\end{eqnarray}}

\newcommand{\ba}{\begin{array}}
\newcommand{\ea}{\end{array}}

\begin{document}
\title{Dimensional dependence of the metal-insulator transition}
\author{Antonio M. Garc\'{\i}a-Garc\'{\i}a}
\affiliation{Physics Department, Princeton University, Princeton,
New Jersey 08544, USA}
\affiliation{The Abdus Salam International Centre for Theoretical
Physics, P.O.B. 586, 34100 Trieste, Italy}
\author{Emilio Cuevas}
\affiliation{Departamento de F\'{\i}sica, Universidad de Murcia,
E-30071 Murcia, Spain}
\date{\today}
 \begin{abstract}
We study the dependence on the spatial dimensionality of different quantities
relevant in the description of the Anderson transition by combining numerical
calculations in a $3 \leq d \leq 6$ disordered tight binding model with theoretical
arguments. Our results indicate that, in agreement with the one parameter scaling
theory, the upper critical dimension for localization is infinity. Typical properties
of the spectral correlations at the Anderson transition such as level repulsion or a
linear number variance are still present in higher dimensions though eigenvalues
correlations get weaker as the dimensionality of the space increases. It is argued
that such a critical behavior can be traced back to the exponential decay of the
two-level correlation function in a certain range of eigenvalue separations. We also
discuss to what extent different effective random matrix models proposed in the
literature to describe the Anderson transition provide an accurate picture of this
phenomenon. Finally, we study the effect of a random flux in our results.
\end{abstract}
\maketitle


\section{Introduction}

After almost fifty years of the landmark paper by Anderson \cite{anderson} about 
localization, the study of the properties of a quantum particle in a random potential
is still one of the central problems of modern condensed matter physics.

In the the early days of localization theory, research was largely focused on the
determination of the critical disorder at which the the metal-insulator transition
--also referred as the Anderson transition-- occurs as a function of the connectivity
of the lattice. In the original Anderson's paper this was achieved by looking at the
limits of applicability of a locator expansion.\cite{anderson,thouless} Later on, a more
refined estimation based on the solution of a self-consistent equation \cite{abu}
provided with a similar answer. The self-consistent method is only exact in the case of
Cayley tree but it is believed to be accurate if the spatial dimensionality is large
enough. We note that in the locator expansion \cite{anderson} the metal-insulator
transition is induced by increasing the hopping amplitude of an initially localized
particle.

In the seventies the application of ideas and techniques of the theory of phase
transitions such as scaling and the renormalization group \cite{wegner} opened new
ways to tackle the localization problem especially in low dimensional systems. These
progress led eventually to the proposal of the one parameter scaling theory \cite{one}
which, despite being still under debate, has become the 'standard' theory of localization.  
In the one parameter scaling theory, localization in a given disordered sample is described
by the dimensionless conductance $g$. This quantity \cite{thouless2} is defined either as
the sensitivity of a given quantum spectrum to a change of boundary conditions in units
of the mean level spacing $\Delta$ or as $g = E_c/\Delta$ where $E_c$, the Thouless energy,
is an energy scale related to the classical diffusion time to cross the sample. The
dimensionless conductance $g$ is sensitive to localization effects. In a metal (insulator),
it increases (decreases) monotonically with the system size, $L$. 

Under the assumption that the dimensionless conductance is only a function of the
system size and by using simple scaling arguments, the one parameter scaling theory
predicts that the metal-insulator transition is characterized by a scale invariant
dimensionless conductance $g =g_c$. The lowest dimension in which the metal-insulator
transition occurs is $d > 2$. In two and lower dimensions destructive interference
caused by backscattering produces exponential localization of the eigenstates in real
space for any amount of disorder in the limit $L \rightarrow \infty$. In this picture,
the Anderson transition is considered as a standard second order transition with critical
exponents $s,\nu$ that control how the conductivity $\sigma \propto |W-W_c|^s$ vanishes
or the localization length $\xi \propto |W_c -W|^{-\nu}$ diverges as the critical disorder
$W_c$ is approached.
 
In $d = 2 + \epsilon$ ($\epsilon \ll 1$) the transition occurs in the weak disorder
region and consequently an analytical treatment is possible. Diagrammatic perturbation
theory and field theory techniques \cite{wegner,voller} predict that $\nu \sim 1/\epsilon$
and $W_c \propto \epsilon$. By contrast, the critical exponent associated to the Cayley
tree, which should be close to that of a disordered conductor in $d \gg 2$ dimensions,
is $\nu = 1/2$.\cite{cayley}  In the context of second order phase transitions this 
value corresponds with the upper critical dimension $d_u$, namely, for $d \geq d_u$
fluctuations are irrelevant and the mean field approximation become exact. For the
localization problem different values $d_u = 4,6,8,\infty$ of the upper critical
dimension have been reported.\cite{sus} The results of this paper discard $d_u=4,6$ and 
indicate that $d_u \to \infty$ is the upper critical dimension. However we would like to
point that the exact significance of the upper critical dimension for localization is
unclear. It is not known what fluctuations are suppressed at the upper critical dimension
and to what extent spectral or transport properties at criticality are affected.

The Anderson transition in a disordered conductor is a consequence of a highly non trivial
interplay  between quantum destructive interference effects and quantum tunneling. In low
dimensions, $d \sim 2$, weak quantum destructive interference effects induce the Anderson
transition. Analytical results are available based on perturbation theory around the
metallic state.\cite{voller,wegner} In high dimensions, $d \gg 2$, quantum tunneling is
dominant and the locator expansion \cite{anderson} or the the self-consistent
formalism \cite{abu} can be utilized to describe the transition. We note that in these
papers corrections due to interference of different paths are neglected.

The progress in numerical calculations during the last twenty years has increased
dramatically our knowledge \cite{numand,sko,schreiber,sch1,ever} of the metal-insulator
transition specially in intermediate dimensions such as $d = 3, 4$ for which a rigorous
analytical treatment is not available. Below we cite a few of its most relevant results.
  
It was verified that for a disorder strength below the critical one, the system has
a mobility edge at a certain energy which separates localized from delocalized
states.\cite{schreiber} Its position moves away from the band center as the disorder
is decreased. Delocalized eigenstates, typical of a metal, are extended through the
sample and the level statistics agree with the random matrix prediction \cite{mehta}
for the appropriate symmetry. The spectral correlations at the Anderson transition,
usually referred to as critical statistics \cite{sko,KMN}, are scale invariant and
intermediate between the prediction for a metal and for an insulator.\cite{sko} By
scale invariant we mean any spectral correlator utilized to describe the spectral
properties of the disordered Hamiltonian does not depend on the system size. 
 
Eigenfunctions at the Anderson transition are multifractals \cite{ever,schreiber}
(for a review see \cite{mirlin,cuevas}), namely, their moments present an anomalous
scaling,
${\cal P}_q=\int d^dr |\psi({\bf r})|^{2q} \propto L^{-D_q(q-1)}$
with respect to the sample size $L$, where $D_q$ is a set of exponents describing the
Anderson transition.  

The main features of the Anderson transition only depend on the dimensionality of the
space and the universality class {\cite{atflux,slevin}}, namely, the presence or not
of a magnetic field (or other time reversal breaking mechanism) or a spin-orbit
interaction. The dependence with the universality class diminishes as the spatial
dimension increases. It has also been reported that certain spectral correlators 
at the Anderson transition are sensitive to different boundary conditions. \cite{mont}

All of these numerical findings are compatible with the one parameter scaling theory.
The applicability of the $\epsilon$-expansion ($d = 2 + \epsilon$) is by contrast much
more restricted. A naive extrapolation to $\epsilon = 1,2$ yields
$\nu_{3D} \sim 1/\epsilon = 1$, $\nu_{4D} = 1/2$ thus suggesting that the upper critical
dimension is four. However numerical calculations \cite{sch1,kramer} show undoubtedly that 
$\nu_{3D} \sim 3/2$ and $\nu_{4D} \sim 1$. Similarly, up to $d = 4$, the self-consistent 
theory overestimate by more than a factor two the value of the critical disorder at which
the Anderson transition occurs. This suggests none of the theories traditionally utilized
to describe the metal-insulator transition can be really extrapolated to the physically
relevant case of $d =3$. In order to make progress in this difficult problem a new basis
for the study of the Anderson transition in any dimension is necessary. In this paper we
have a more modest goal: a detailed exploration of the dependence of different quantities
defining the Anderson transition on the spatial dimensionality.
  
We propose simple relations that describe how the parameters defining the Anderson
transition depend on the dimensionality of the space. It is argued that the upper critical
dimension must be infinite. Our results are supported by numerical evidence from a
disordered Anderson model in a hyper-cubic lattice in $3 \leq d \leq 6$. This is the first
time than the Anderson transition in $d=5,6$ is investigated numerically in the literature
(for some recent results in a small asymmetric lattice in $d=5$ we refer to
Ref. \cite{markos}).

The organization of the paper is as follows. In section II, we introduce the model to
be studied, explain the technical details of the numerical simulation, locate the mobility
edge in different dimensions and investigate how the critical exponent and the critical
disorder depend on the spatial dimensionality. In Sec. III, spectral correlation at the
Anderson transition are investigated, three different regions of the two level correlation
function are distinguished, we also study the dependence of the slope of the number variance
and the asymptotic decay of the level spacing distribution with the spatial dimensionality.
It is also discussed the range of validity of certain phenomenological models commonly used
in the literature to describe the spectral correlations at the Anderson transition. Finally,
we examine the effect of a magnetic flux in our results.

\section{Critical disorder, critical exponents and upper critical dimension}

In this section we determine the critical disorder and critical exponents for different 
dimensions and then discuss their dependence with the spatial dimensionality.

\subsection{The model: Technical details}
Our starting point is the standard tight-binding Anderson model on a hyper-cubic
$L^d$ lattice with $d = 3, \dots , 6$
\begin{equation}
\label{ourmodel}
{\cal H}=\sum_i \epsilon_i a^{\dag}_i a_i + \sum_{ij} t_{ij} a^{\dag}_i a_j\;,
\end{equation}
where the operator $a_i (a^{\dag}_i)$ destroys (creates) an electron at the $i$th site
of the lattice and $t_{ij}$ is the hopping integral between sites $i$ and $j$ which is
non zero only for nearest neighbors. In the following we take $t_{ij} = 1$ for $i,j$
and the lattice constant equal to unity, which sets the energy and length units,
respectively.  The uncorrelated random energies $\epsilon_i$ are distributed with constant
probability within the interval $(-W/2, W/2)$, where $W$ denotes the strength of the
disorder and hard-wall boundary conditions are imposed in all directions.

In order to proceed we compute eigenvalues of the Hamiltonian Eq. (\ref{ourmodel}) for
different volumes and disorders by using techniques for large sparse matrices, in
particular a Lanczos tridiagonalizaton without reorthogonalization method.\cite{CW85}
We restrict ourselves to a small energy window $(-2, 2)$ around the center of the band.
Calculations have been carried out in samples of sizes up to $L=30$ for $d=3$, $12$ ($d=4$),
$10$ ($d=5$) and $7$ ($d=6$). The number of random realizations is such that for a given
triad of $\{ d, L, W \}$ the number of eigenvalues obtained is at least $3 \times 10^5$.
In order to study the level statistics around the mobility edge more accurately, this
number was increased to $20 \times 10^6$ at the critical disorder.

Eigenvalues thus obtained are appropriately unfolded, i.e, they were rescaled so that the
spectral density on a spectral window comprising several level spacings is unity. 

\subsection{Location of $W_c(d)$ and $\nu(d)$}
Our first task is to find out the critical disorder $W_c$ and the critical exponent
$\nu$ for different dimensions in the small spectral window $(-2,2)$ around to the origin.  
In order to proceed we determine the location of the mobility edge close to the band center 
by using the finite size scaling method. \cite{sko} First we evaluate a certain spectral
correlator for different sizes $L$ and disorder strengths $W$. Then we locate the mobility
edge by finding the disorder $W_c$ such that the spectral correlator analyzed becomes
size-independent. In our case we investigate the level spacing distribution $P(s)$ 
(probability of finding two neighboring eigenvalues at a distance 
$s = (\epsilon_{i+1} - \epsilon_{i})/\Delta $, with $\Delta$ being the local mean
level spacing). The scaling behavior of $P(s)$ is examined through the following function
of its variance\cite{C99}
\begin{equation}
\label{scaling}
\eta (L,W) = [{\rm var} (s)- {\rm var_{WD}}\,] / [\,{\rm var_{P}}-{\rm var_{WD}}] \;,
\label{eta}
\end{equation}
which describes the relative deviation of ${\rm var} (s)$ from the Wigner-Dyson (WD)
limit. In Eq. (\ref{eta}) ${\rm var} (s)=\langle s^2\rangle-\langle s \rangle^2$, where
$\langle \dots \rangle$ denotes spectral and ensemble averaging,
and ${\rm var_{WD}} = 0.286$ and ${\rm var_P}=1$ are the variances of WD and Poisson
statistics, respectively. Hence $\eta = 1 (0)$ for an insulator (metal). Any other
intermediate value of $\eta$ in the thermodynamic limit is an indication of a mobility
edge. 

\begin{figure}
\includegraphics[width=1.0\columnwidth,clip]{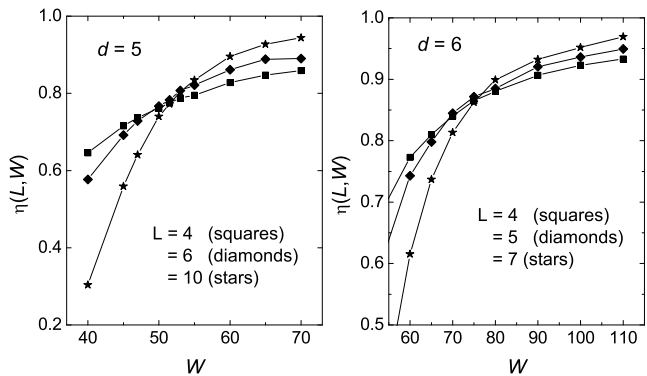}
\caption{Scaling variable $\eta$ as a function of disorder $W$ for different
volumes in $d=5$ (left panel) and $d=6$ (right panel). Energies in the interval
$(-2, 2)$ have been considered. The system undergoes an Anderson transition at
$W_c = 51.4$ and $74.5$ for $d= 5$ and $6$, respectively (see text).}  
\label{fig29} 
\end{figure}

In Fig. 1 we plot the $W$ dependence of $\eta$ for different system sizes in $d =5$
(left panel) and $d=6$ (right panel). The critical disorder $W = W_c$ signaling the
Anderson transition corresponds with the point for which $\eta$ is independent of $L$.
For a weaker (stronger) disorder, $\eta$ tends to the metallic (insulator) prediction. 
This is the first time that an Anderson transition is found in such a high dimensional
disordered system. For a precise determination of the critical disorder $W_c$ and the
critical exponent $\nu$ we look at the correlation length near $W_c$
\begin{equation}
\xi(W) = \xi_0 |W-W_c|^{-\nu}\;, \label{xiw}
\end{equation}
where $\xi_0$ is a constant. The numerical values of $W_c$ and $\nu$ are obtained 
by expressing $\eta(L,W)=f[L/\xi(W)]$ and then performing an expansion around the
critical point
\begin{equation}
\eta(L,W) = \eta_c+ \sum_{n}C_n(W-W_c)^nL^{n/\nu}\;. \label{fit}
\label{alfa1}
\end{equation}
In practice, we have truncated the series at $n=4$. For each dimension ($d = 5, 6$) we
have performed a statistical analysis of the data in the windows shown in Fig. 1 with
the Levenberg-Marquardt method for nonlinear least-squares models. The most likely fit
is determined by minimizing the $\chi^2$ statistics of the fitting function (\ref{fit}).
We found the following critical disorders $W_c = 51.4 \pm 0.4$ in $d=5$ and
$W_c = 74.5 \pm 0.7$ in $d=6$, and the corresponding critical exponents are equal to
$\nu=0.84 \pm 0.06$ and $\nu=0.78 \pm 0.06$, respectively. A similar analysis for the
$d=3$ and $d=4$ systems results in $W_c = 15.22 \pm 0.08$ and $\nu=1.52 \pm 0.06$ for
$d=3$, and $W_c = 29.8 \pm 0.2$ and $\nu=1.03 \pm 0.07$ for $d=4$. We note that in the
$d=3$ case, the deviation of $W_c$ from the accepted value $W_c \sim 16.5$ is due to the
utilization of rigid boundary conditions. See Fig. 2 for a plot of $W_c$ and $\nu$ as a
function of the spatial dimensionality.

\subsection{Theoretical analysis of $W_c(d)$ and $\nu(d)$}
In certain limiting cases $W_c$ and $\nu$ are known analytically. For instance, if
effects related to interference among different paths are neglected  \cite{abu}, the
standard tight binding Anderson model is effectively defined on a  Cayley tree and
$\nu = 1/2$.  On the other hand if only interference corrections to the metallic limit
are considered then $\nu = 1/(d-2)$.\cite{voller} The former prediction is supposed to
be approximately valid  for $ d \gg 2$ and the latter for $d =2 +\epsilon$ and
$\epsilon \ll 1$. From the above numerical results it is clear that none of these limits
is appropriate in the range of intermediate dimensions of interest. Additionally, it is
believed \cite{abu} that corrections to the $\nu = 1/2$ result should go as $\sim 1/d$
since this is the dependence on the spatial dimensionality of the neglected diagrams
describing interference effects. Combining these two facts we propose that
\be
\nu = \frac{1}{d-2} + \frac{1}{2}
\ee
for all dimensions. As is shown in Fig. 2 (right panel), this relation verifies all 
limiting cases and reproduce the numerical results accurately. According to the above
relation, the upper critical dimension for localization is infinite. This result is
fully supported by the analysis of the spectral correlations (see next section).
Moreover, in a recent paper \cite{michael} it has been proved rigorously that the level
statistics of a disordered system in a Cayley tree ($\nu = 1/2$) is Poisson as for an
insulator. As was mentioned previously, the Cayley tree represents to a $d$-dimensional
conductor where all interference effects between different paths are neglected. It is
thus supposed to be an accurate description of a disordered conductor only in the limit
$d \gg 2$. 

On the other hand, if the one parameter scaling theory is valid, quantum diffusion never
stops (see next section) for any finite dimension. Level repulsion typical of a metal
will be present in any finite dimension so level statistics at criticality can be that
of an insulator only in the $d \rightarrow \infty $ limit.  But this precisely the result
for the Cayley tree  \cite{michael} which corresponds with the upper critical dimension
for localization. It is thus clear that the upper critical dimension must be infinity. 

A similar analysis can be carried out for the critical disorder $W_c$. The original
estimation of Anderson \cite{anderson}, $W_c = 4 K \ln(W_c/2)$ (where $K$ is the
connective constant which is a little bit less than the number of nearest neighbors
minus one) greatly overestimates $W_c$. This is hardly surprising since the Anderson's
calculation involves crude approximations and consequently should be considered as an
order of magnitude estimation rather than an accurate prediction. For instance, deviations
coming from interference effects are neglected in this scheme. Roughly speaking they tend
to reduce $W_c$ by an amount of order $W_c/d$.

In the opposite limit, $d = 2 +\epsilon$, simple perturbation theory \cite{voller}
yields $W_c \propto d-2$. The discrepancy observed with the analytical results in
the limit of high dimensionality prevent us from proposing an interpolating relation
as in the case of the critical exponent. However we have noticed that a much better
agreement with the numerical results is achieved if an effective connective constant
$K_{{\rm eff}} = K/2$ is utilized (solid line in left panel of Fig. 2).
Furthermore, the remaining deviation gets smaller as the spatial dimensionality increases 
thus suggesting that it may be produced by destructive interference effects ($\sim W_c/d$).

\begin{figure}
\includegraphics[width=1.0\columnwidth,clip]{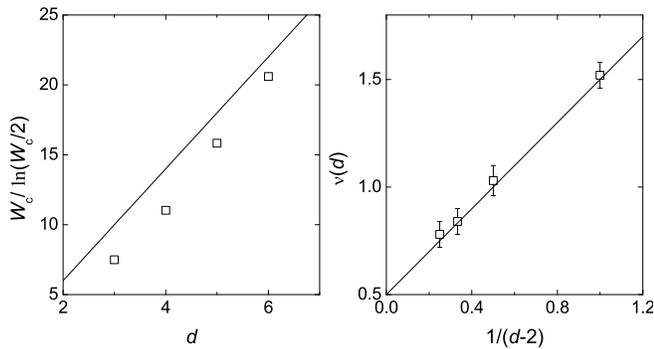}
\caption{$W_c$ (left panel) and $\nu$ (right panel) as a function of the dimensionality
$d$. The simple interpolating formula (solid line) Eq. (5) describes quite well the
numerical results for the critical exponents $\nu$.} 
\label{fig2} 
\end{figure}
 
\section{Level statistics}
In this section we investigate the level statistics at the Anderson transition. 
We shall mainly focus on its dependence on spatial dimensionality and the exact
functional form of the two-level correlation function (TLCF).

\subsection{Theoretical analysis of $R_2(s)$}

Our starting point is the connected TLCF, 
\be
\label{r2s1}
R_{2}(s)={\frac{1}{\langle\rho(\epsilon)\rangle^2}}
\langle \rho(\epsilon -\omega/2)\rho(\epsilon +\omega/2) \rangle \;,
\ee
where $\rho(\epsilon)$ is the density of states at energy $\epsilon$,
$\langle \; \rangle $ denotes averaging over disorder realizations 
and  $s=\omega/\Delta$ where  $\Delta=1/L^d\langle\rho(\epsilon)\rangle$ 
in the mean level spacing. Once the spectrum has been unfolded $R_2(s)$ can
be simply written as
\begin{equation}
\label{r2s2}
R_2(s)=\delta(s)+\sum_n p(n;s)\;,
\end{equation}
where $p(n;s)$ is the distribution of distances $s_n$ between $n$ other energy levels
and $\delta(s)$ describes self-correlation of levels.\cite{mehta} In numerical
computations we use Eq. (\ref{r2s2}) since it gives much more accurate results than
Eq. (\ref{r2s1}).

According to the one parameter scaling theory, the spectral properties depend on the
dimensionless conductance $g$ which is a function of the system size $L$ only.
In a metal $g \rightarrow \infty$ for $L \rightarrow \infty$, the Hamiltonian can be
accurately approximated by a random matrix with the appropriate symmetry and Wigner-Dyson
statistics applies.\cite{mehta} For instance, for broken time reversal invariance, 
$R_2(s)=\delta(s)+1-\frac{\sin^2(\pi s)}{\pi^2s^2}$. In an insulator, eigenvalues 
uncorrelated, Poisson statistics applies and $R_2(s) = \delta(s)$.

Right at the Anderson transition, the dimensionless conductance $g = g_c$ is size
independent and level statistics are supposed to be universal and intermediate between
Wigner-Dyson and Poisson statistics. Unfortunately there are few analytical results for
the TLCF at criticality. In the $d \gtrsim 2$ region \cite{wegner},
$g_c \sim 1/(d-2) \gg 1$. The TLCF, can only evaluated explicitly in the limit
$\rightarrow 2$ \cite{andre} and $g \gg 1$ where $R_2(s) \sim s$ (time-reversal),
for $s \ll g_c$, as for a metal, and $R_2(s) \sim {\rm e}^{- A s/g_c}$ for $s \gg g_c$,
with $A$ being a factor of order unity. The Anderson transition is thus characterized
by level repulsion combined with an exponential decay of the TLCF.

In higher dimensions the exact form of the TLCF is not known. However, we note that the
scale invariance of the spectral correlations at the Anderson transition restricts the
decay of the TLCF in the $s \gg g_c$ region to be power-law or exponential\cite{conformal}. 
Our numerical results (see Fig. 3) for $d \geq 3$ support also this picture.

The limit of long times and small energy differences $s \ll g_c$, $s \ll g_c$ is well
understood in high dimensions as well. Level repulsion of neighboring eigenvalues
$R_2(s) \propto s$, typical of a metal, should be a generic feature in any dimension. 
According to the one parameter scaling theory, the averaged moments of the particle
position at the Anderson transition increase asymptotically $t \rightarrow \infty$ as
$\langle r(t)^{2m}\rangle \sim t^{2m/d}$
where $m$ is a positive integer. As the spatial dimensionality  $d$ increases the
diffusion is slowed down but it never stops even for long times. This is an indication
that the spectral correlations for sufficiently small energy intervals are similar to
those of a metal and, as a consequence, $R_2(s) \sim s$ for $s \ll g_c$ (see Fig. 3).                                                                                                                                                                                                                                                                                                          
                                                           
Finite size effects modify the TLCF in the critical region.\cite{aronov} In any finite
system at criticality the localization length $\xi \propto |E_c - E|^{-\nu}$ ($E_c$ is
the location in energies of the mobility edge) is finite and the dimensionless conductance
is not, strictly speaking, scale invariant, $g(L_{\xi}) = g_c[1+(L_{\xi}/L)^{1/\nu}]$
where $L_{\xi}$ is the localization length for a given $E \sim E_c$.
As a consequence \cite{aronov}, the TLCF develops a power-law tail,
$R_2^{{\text tail}}(s) \propto s^{\gamma -2}$ with  $\gamma = 1 - 1/(\nu d)$ for
$s > \Delta_{\xi}/\Delta$, where $\Delta_{\xi}$  is the mean level spacing in a
localization volume $\xi^d$. This tail is not related with the properties of the
critical point but rather with how the system approaches to it. In $d = 2+\epsilon$,
$\nu = 1/\epsilon \gg 1$ and $R_2^{{\text tail}}(s) \sim 1/s$.

As a summary, we can distinguish three different regions in the critical TLCF, 
for $s \ll g_c$,  $R_2(s) \propto s$, for $s \gg g_c$, $R_2(s)$ decays exponentially.  
For  $s > \Delta_{\xi}/\Delta$ decays as power-law due to finite size effects. In order
to observe the exponential decay related to the critical point our system size must be
such that $g_c  >   \Delta_{\xi}/\Delta$. Finally we note that the exact dependence of
$g_c$ on  the spatial dimensionality it is not known. We are only aware of the prediction
of Vollhardt and Wolfle \cite{voller} by using a self-consistent diagrammatic theory valid
for $ 4 > d > 2$, $g_c(d) = c_d/(d-2)$ with $c_d = (2/\pi)[S_d/(2\pi)^d]$ and $S_d$ the
surface of a $d$-dimensional sphere of radius unity. In principle it should be accurate
only for $d \gtrsim 2$ though it is unclear its exact range of validity.

\subsection{Numerical analysis of $R_2(s)$}
After the theoretical analysis we are now ready to present our numerical results for
$R_2(s)$ at the Anderson transition in $d = 3 - 6$ dimensions. Our motivation is to
study the existence and extension of the three regions introduced above: level repulsion,
power-law and exponential decay. Indeed our numerical results clearly show these three
regimes in all dimensions $d = 3 - 6$ investigated.

We have first verified (not shown) that for sufficiently large $s$,
$R_2(s) \sim 1/s^{\gamma}$. The numerical value of the exponent $\gamma$ was in full
agreement with the theoretical prediction $\gamma = 1- (\nu d)^{-1}$. 

Then we investigate to what extent level repulsion typical of the Anderson transition
in $d=3,4$ is still present in higher dimensions. As is observed in Fig. 3, left, for
sufficiently small $s$, $R_2(s) \sim s$ for all dimensions studied. The solid lines
lines are linear fits of the form $R_2(s)=C+Ds$ with fitting parameters $D=6.6 \pm 0.8$
for $d=3$, $15.0 \pm 1.2$ ($d=4)$, $101 \pm 5$ ($d=5)$, and $373 \pm 32$ ($d=6$). The
parameter $C$ is equal to zero within the error bars in all cases. This is consistent
with the prediction of the one parameter scaling theory that quantum diffusion never
stops. However the range in which level repulsion is observed decreases dramatically
with the spatial dimensionality thus suggesting that the critical conductance $g_c$
also decreases rapidly with the dimension. It is hard to give a more quantitative
prediction of $g_c$ as a function of the spatial dimensionality: the estimation of
Vollhardt and Wolfle\cite{voller} mentioned previously fails for $d > 3$. 
Another option is to extrapolate the result in the diffusive regime \cite{mirkra}
$R_2(s) \sim s(1+a_d/g^2)$ for $s \ll g_c$ to the critical one. However the geometrical
coefficient $a_d$ diverges also for $d > 3$.

Our numerical results (see Fig. 3, right) show that for larger spectral separations
$s \geq g_c$, the linear repulsion is replaced by an exponential decay. The solid lines
correspond to a linear fit $\ln [1-R_2(s)]= C - Ds$ with fitting parameters
$D=3.7 \pm 0.1$ for $d=3$, 
$4.7 \pm 0.1$ ($d=4)$,
$5.6 \pm 0.5$ ($d=5)$, and
$9.5 \pm 0.2$ ($d=6)$. 
The maximum value of $s$ plotted was chosen attending to technical criteria. For larger
values of $s$, $1-R_2(s)$ fluctuates around zero thus suggesting that the maximum precision
of the computer has been reached.

We note that such exponential decay has already observed in certain one-dimensional
disordered systems with long range hopping \cite{ever} and in phenomenological short-range
plasma models \cite{bogo3} whose spectral properties are strikingly similar to those of a
disordered system with short range hopping at the Anderson transition. In these
one-dimensional systems it can be proved analytically that
$R_c(s) \sim {\text e}^{-A s/g}$ where $A$ is a constant of order unity. It is thus
tempting to speculate that in our case $g_c \sim 1/D$ with $D$ the fitting parameter above.
However a deeper analytical knowledge about the Anderson transition is needed to discard
that additional geometrical factors (as $a_d$ above) enter in the exponent of $R_2(s)$
making thus less evident the relation between $g_c$ and $D$.

\begin{figure}
\includegraphics[width=1.0\columnwidth,clip]{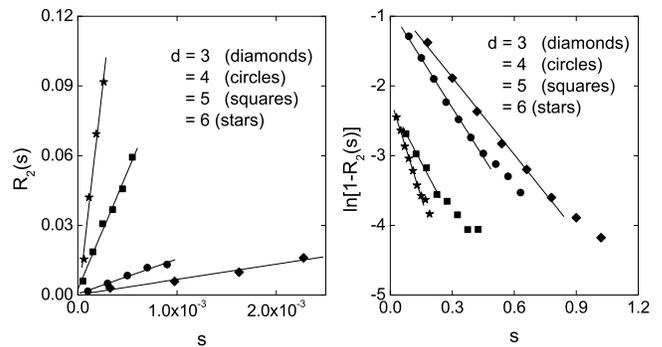}
\caption{$R_2(s)$ for $d=3-6$ at the Anderson transition for different $s$ ranges. In
the left panel we look at the region of small $s$ where level repulsion is still observed. 
As the dimensionality increases the Anderson transition occurs for stronger disorder and
the region of level repulsion is smaller. In the right panel it is shown the window of $s$
in which exponential decay is observed. Such decay is responsible of typical features of
the Anderson transition as a linear number variance or a scale invariance spectrum.
For the sake of clarity we have removed the power-law contribution
$R_2(s) \propto 1/s^{\gamma}$. It is well established that this term does not really
describe the properties at the Anderson transition but rather how the system approaches
to it. Moreover its contribution to the number variance and other spectral correlators
is negligible with respect to the exponential contribution.}
\label{fig3} 
\end{figure}

\subsection{Spectral correlators}
Level statistics at the Anderson transition are usually investigated by computing certain
spectral correlators from the TLCF or higher $n$-level correlation functions. The level
spacing distribution $P(s)$ is a popular choice to study the correlations of eigenvalues 
separated short distances of order the mean level spacing. On the other hand, the number
variance $\Sigma^2(\ell)=\langle (N_\ell -\langle N_\ell \rangle)^2 \rangle$  
($N_\ell$ is the number of eigenvalues in an interval of length $\ell$) provides useful
information about spectral correlations for distances much larger than the mean level
spacing. 

Numerical calculations in $d \leq 4$ at the Anderson transition have found that 
$P(s) \rightarrow 0$ for $s \rightarrow 0$, as in a metal. However the number variance
is asymptotically linear $\Sigma^2(\ell) \sim  \chi \ell$,  as in an insulator but with
a slope $\chi < 1$. The origin of this linear behavior can be explained heuristically
\cite{chi} by using the one parameter scaling theory and making the plausible approximation
that eigenvalues interact only if their separation (in units of the mean level spacing) is
smaller than $g_c$. In the critical region it is also expected that
${P}(s) \sim {\rm e}^{-As}$ ($A > 1$) for $s \gg 1$ similar to the insulator limit
${P}(s) = {\rm e}^{-s}$.

A natural question to ask is whether these spectral features also holds at the Anderson
transition in higher dimensions $d = 5, 6$. Our numerical results (see Fig. 4) fully
confirm that both $P(s)$ and $\Sigma^2(\ell)$ have all the signatures of critical
statistics. The plots of $P(s)$ and $\Sigma^2(\ell)$ correspond to the maximum $L$ used
in each dimension though almost identical results are obtained for smaller volumes (not
shown). The straight lines in Fig. 4 are fits of the form  
$\Sigma^2(\ell)= C + \chi \ell$ and $\ln P(s) = D -A s$. The best fitting parameters
$\chi$ and $A$ are plotted in Fig. 5 as a function of the spatial dimensionality. It is
clearly observed that the slope of the number variance $\chi$ increases and $A$ 
decreases with the spatial dimensionality but does not reach the Poisson limit
$\chi = A = 1$. This confirms that the upper critical dimension must be $d_u > 6$ and
strongly suggests that it is indeed infinity as this is, according to the fitting in
Fig. 5, the dimension in which $\chi = A = 1$.

\begin{figure}
\includegraphics[width=1.0\columnwidth,clip]{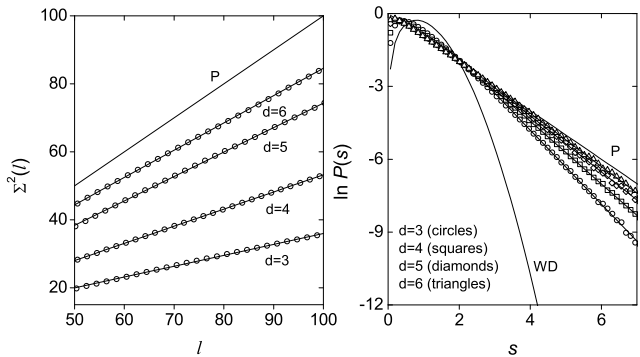}
\caption{Number variance $\Sigma^2(l)$ (left panel) and $P(s)$ (right panel) at
the Anderson transition for $d=3-6$. The system tends to the Poisson limit (P)
as the dimension is increased. WD denotes the Wigner-Dyson distribution.}  
\label{fig4} 
\end{figure}

We are especially interested on the specific dependence of $\chi$ and $A$  with 
the spatial dimensionality. In $d = 2 + \epsilon$ the Anderson transition occurs in
the weak disorder region, $g_c \gg 1$ and $\chi \sim 1/g_c \sim d -2 \ll 1$. On the
other hand, the prediction for $d \rightarrow \infty$ (Cayley tree) is
$A = \chi = 1$.\cite{michael} In principle, corrections to the Cayley  
tree limit due to interference between different paths decay as $1/d$ or faster
so it is tempting to conjecture that $\chi =1 - C/(d-2)$ and $A=1 + D/(d-2)$.
The numerical results of Fig. 5 confirm this dependence especially for the parameter $A$.
In the case of $\chi$ the situation is less clear. A reason for the discrepancy with the
theoretical prediction could be that $d \sim 3$ is still far from the limit $d \gg 2$ in
which  $\chi =1 - C/(d-2)$ holds. Indeed we have observed that our numerical data are
better described (dotted line in Fig. 5) by $\chi = \tanh[C\,(d-2)]$ with
$C = 0.29 \sim 1/\pi$. Such a dependence of $\chi$ on hyperbolic functions has already
been reported on the generalized random matrix models\cite{ever} whose spectral
correlations are strikingly similar to the ones at the Anderson transition.  

The straight lines in Fig. 5 are linear fits to the conjectured relations with fitting
parameters $C=0.78 \pm 0.06$ and $D=0.55 \pm 0.01$. From a physical point of view these
numerical results are a further confirmation that analytical approaches to the Anderson
transition starting from the metallic limit and adding interference corrections or  
starting from the insulator state and inducing the transition to a metal by increasing
the tunneling amplitude fail to capture key features of the Anderson transition in 
intermediate dimensions where both mechanisms are at work. 

\begin{figure}
\includegraphics[width=1.0\columnwidth,clip]{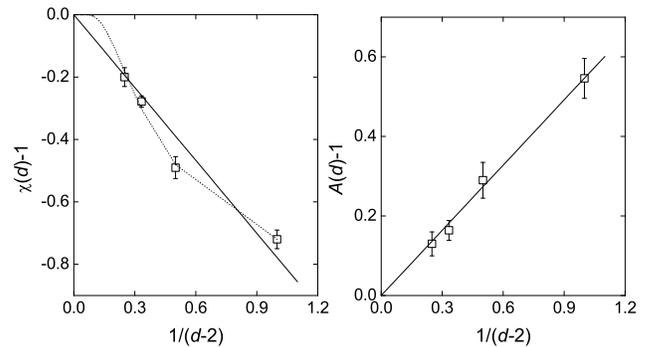}
\caption{Dependence of the slope of the number variance $\chi$ (left panel) and
the asymptotic decay of the level spacing distribution $A$ (right panel) with the
dimensionality of the space.}  
\label{fig5} 
\end{figure}

\subsection{Random matrix models and the Anderson transition}
Typical signatures of critical statistics have also been found in both 
generalized random matrix models \cite{Moshe,ant4,ever} whose joint distribution of 
eigenvalues can be mapped onto the Calogero-Sutherland model at finite temperature 
and phenomenological short-range plasma models whose joint distribution
of eigenvalues  {\cite{bogo3}} is given by the classical Dyson gas with the logarithmic
pairwise interaction restricted to a finite number $k$ of nearest neighbors (the spectral
correlations of this model are usually referred to as Semi-Poisson statistics though this
name refers to the case $k =2$). In the latter explicit analytical solutions for all
correlation functions are available for general $k$. Although these models reproduce
typical properties of critical statistics such as spectral scale invariance,  level
repulsion and linear number variance, they are quantitatively different. In the generalized
matrix models the joint distribution of eigenvalues can be considered as an ensemble of
free particles at finite temperature with a nontrivial statistical interaction. The
statistical interaction resembles the Vandermonde determinant and the effect of a finite
temperature is to suppress smoothly correlations of distant eigenvalues. In the case of
the short range plasma model \cite{bogo3} this suppression is abrupt since only nearest
neighbor levels interact each other. A natural question to ask is which of those mechanisms
is dominant in the Anderson transition studied in this paper. We have found a method to
distinguish between them. In the short range plasma model $A \chi = 1$ ($A$ describes the
exponential decay of $P(s) \sim {\text{e}}^{-As}$). By contrast, in the generalized random
matrix models $A \chi$ falls between $1/2$ in the region of weak disorder to unity in
the region for strong disorder. On the other hand in our case -- a disordered tight binding
model at the Anderson transition -- $A \chi$ ranges from $0.44$ in $d=3$ to $0.9$ in
$d=6$ in agreement with the prediction of the generalized random matrix models. Our results
thus suggest that the abrupt suppression of spectral correlations typical of Semi-Poisson
statistics can describe the spectral correlation at the Anderson transition in $d \gg 2$
but not for intermediate dimensions. 

\subsection{Effect of a magnetic flux}
So far all results we have presented correspond to the case of time-reversal invariance.
We have also investigated the effect of a random flux at criticality in $d=3-6$. This has
been achieved by the substitution
$t_{ij} \to t_{ij}{\text{e}}^{\text{i}\theta_{ij}}$ in the Hamiltonian Eq. (1).
The phases $\theta_{ij}$ were chosen to be uniformly distributed in the interval
$[-\pi,\pi]$. In $d =3$, in agreement with previous claims in the literature\cite{atflux}, 
small differences with respect to the time reversal invariance case were found in $W_c$
and in $P(s)$ in the $s \ll 1$ limit. Typically these effects are related with weak
localization like corrections that are strongly affected by the flux.  However in
$d= 5, 6$ the time-reversal and the time-broken cases were almost indistinguishable.
This suggests that the mechanism of localization leading to weak localization corrections
based on destructive interference is less important in $d \gg 2$ dimensions.


\section{Conclusions}
We have studied the dependence on the spatial dimensionality of different quantities
relevant in the description of the Anderson transition. As a result we have concluded
that the upper critical dimension for localization is infinity. The level statistics
tend to Poisson statistics, typical of an insulator, as the upper critical dimensionality
is approached. We have also proposed that the exponential decay of the TLCF observed in
numerical calculations is a signature of an Anderson transition. Neither the self-consistent
theory of localization exact in the Cayley tree nor the $\epsilon$-expansion formalism are
accurate for intermediate dimensions. A new basis for the localization problem is thus
called for. Finally, the effect of a magnetic flux and the validity of certain effective
models to describe the spectral correlations at the Anderson transition have been discussed.

\begin{acknowledgments}
A.M.G. was supported by a Marie Curie Action, contract
MOIF-CT-2005-007300. E.C. thanks the FEDER and the Spanish DGI for
financial support through Project No. FIS2004-03117.
\end{acknowledgments}


\end{document}